\documentstyle[aps,epsfig,preprint]{revtex}

\begin{document}

\title{Complete relativistic equation of state for neutron stars}

\author{H. Shen\footnote{Electronic address: songtc@public.tpt.tj.cn}} 

\address{
CCAST (World Laboratory), P.O. Box 8730, Beijing 100080, China\\
Department of Physics, Nankai University, Tianjin 300071, China\footnote{Mailing address}\\
Center of Theoretical Nuclear Physics, National Laboratory of 
Heavy Ion Accelerator, Lanzhou 730000, China\\
Institute of Theoretical Physics, Beijing 100080, China\\
}

\maketitle

\begin{abstract}
We construct the equation of state (EOS) in a wide density range
for neutron stars using the relativistic mean field theory.
The properties of neutron star matter with both uniform and 
non-uniform distributions are studied consistently.
The inclusion of hyperons considerably softens the EOS at high densities.
The Thomas-Fermi approximation is used to describe the non-uniform
matter, which is composed of a lattice of heavy nuclei. 
The phase transition from uniform matter to non-uniform
matter occurs around $0.06\ \rm{fm^{-3}}$, and the free neutrons drip
out of nuclei at about $2.4 \times 10^{-4}\ \rm{fm^{-3}}$.
We apply the resulting EOS to investigate the neutron star properties
such as maximum mass and composition of neutron stars. 

\vspace*{0.5cm}


\end{abstract}

\section{Introduction}
The properties of neutron stars are mainly determined by 
the equation of state (EOS) of neutron star matter, 
which is charge neutral matter in $\beta$-equilibrium.
A comprehensive description of neutron stars 
should include not only the interior region but also the inner 
and outer crusts, therefore, the EOS for neutron stars is 
required to cover a wide density range. 
For the EOS at high densities, there are many efforts 
based on both non-relativistic and relativistic approaches,
which discussed several possible mechanisms to soften the EOS 
at high densities, e.q., by hyperons, kaon condensates, 
or even quark phases~\cite{LA97,HH00,QMC99}.  
When the density lower to $10^{14}\ \rm{g/cm^3}$, 
some heavy nuclei may be formed and matter becomes inhomogeneous. 
There are a few works based on non-relativistic models
describing the EOS at low densities where the heavy nuclei 
exist~\cite{BB71,NV73,PR95}.
Most studies of neutron stars are using the composite EOS,
which is constructed by connecting the EOS at high densities 
to the one at low densities~\cite{GW92,SA20,GR20}. 
Even though the EOS at high densities are based on various 
relativistic many body theories, it has to be combined with 
some non-relativistic EOS at low densities.
The differences in the models used in the different density ranges 
usually lead to some discontinuity and inconsistency
in the composite EOS. 
Therefore, it is very interesting to construct the EOS 
in the whole density range within the same framework.

In this paper, we provide a complete relativistic 
EOS for the studies of neutron stars, 
which is based on the relativistic mean field (RMF)
theory. The RMF theory has been quite successfully and 
widely used for the description of nuclear matter 
and finite nuclei~\cite{SW86,PR90,HS96}.
We study the properties of dense matter with both uniform 
and non-uniform distributions in the RMF framework adopting 
the parameter set TM1, which is known to provide excellent 
properties of the ground states of heavy nuclei including 
unstable nuclei~\cite{ST94}. 
The RMF theory with the TM1 parameter set was also shown 
to reproduce satisfactory agreement with experimental data
in the studies of the nuclei with deformed configuration 
and the giant resonances within the RPA formalism~\cite{HTT95,MT97,MA01}.
At high densities, hyperons may appear as new degrees of freedom
through the weak interaction, the neutron star matter is 
then composed of neutrons, protons, hyperons, electrons, 
and muons in $\beta$-equilibrium.
For the non-uniform matter at low densities,
we perform the Thomas-Fermi calculation, in which 
the RMF results are taken as its input. 
The non-uniform matter is assumed to be composed of a lattice 
of spherical nuclei immersed in an electron gas
with (or without) free neutrons dripping out of nuclei~\cite{O93,SO95}. 
The optimal state at each density is determined by minimizing
the energy density with respect to the independent parameters 
in the model. 
The phase transition from non-uniform matter to uniform matter 
takes place around $10^{14}\ \rm{g/cm^3}$.
The same method (but without the inclusion of hyperons) 
has been used to work out 
the equation of state at finite temperature with various proton 
fractions for the use of supernova simulations~\cite{npa98,ptp98}.

This paper is arranged as follows. In Sec. II, we briefly describe
the RMF theory and its parameters. In Sec. III, we explain the 
Thomas-Fermi approximation used for the description of non-uniform
matter. The resulting EOS in the whole density range is shown and 
discussed in Sec. IV. We apply the relativistic EOS to 
study the constitution and structure of neutron stars in Sec. V.
The conclusion is presented in Sec. VI. 
      
\section{ Relativistic mean field theory}

We briefly explain the RMF theory used to describe the uniform matter.
In the RMF theory, baryons interact via the exchange of mesons.
The baryons considered in the present calculation include
nucleons ($n$ and $p$) and hyperons 
($\Lambda$, $\Sigma$, $\Xi$).
The exchanged mesons consist of isoscalar scalar and vector mesons 
($\sigma$ and $\omega$), isovector vector meson ($\rho$),
and two strange mesons ($\sigma^*$ and $\phi$) which couple only
to hyperons.    
The total Lagrangian density of neutron star matter,
in the mean field approximation, can be written as 
\begin{eqnarray}
\label{eq:L}
{\cal L} &=& \sum_B
\bar\psi_B\left[ i\gamma_\mu\partial^\mu-
\left( m_B-g_{\sigma B}\sigma-g_{\sigma^* B}\sigma^* \right) 
-\left( g_{\omega B}\omega + g_{\phi B}\phi + 
g_{\rho B}\tau_3\rho \right)\gamma^0
\right] \psi_B  
\nonumber\\
 & & 
-\frac{1}{2} m_\sigma^2\sigma^2 
+\frac{1}{3} g_2\sigma^3 
-\frac{1}{4} g_3\sigma^4
+\frac{1}{2} m_\omega^2\omega^2 
+\frac{1}{4} c_3\omega^4
+\frac{1}{2} m_\rho^2\rho^2
\nonumber\\
 & &
-\frac{1}{2} m_{\sigma^*}^2{\sigma^*}^2
+\frac{1}{2} m_\phi^2\phi^2 
+\sum_l\bar\psi_l\left( i\gamma_\mu\partial^\mu - m_l\right) \psi_l \ ,
\end{eqnarray}
where the sum on $B$ is over all the charge states of the baryon 
octet ($p$, $n$, $\Lambda$, $\Sigma^+$, $\Sigma^0$, $\Sigma^-$,
$\Xi^0$, $\Xi^-$), and the sum on $l$ is over the electrons and 
muons ($e^-$ and $\mu^-$). The meson mean fields are denoted as
$\sigma$, $\omega$, $\rho$, $\sigma^*$, and $\phi$.
The inclusion of the non-linear $\sigma$ and $\omega$ terms 
is essential to reproduce the feature of the relativistic 
Brueckner-Hartree-Fock theory and satisfactory properties of
finite nuclei~\cite{ST94}.  
We adopt the TM1 parameter set for the meson-nucleon couplings
and the self-coupling constants and some masses,
which was determined in Ref.~\cite{ST94} by reproducing
the properties of finite nuclei in the wide mass range 
in the periodic table including neutron-rich nuclei.
The RMF theory with the TM1 parameter set was also shown 
to reproduce satisfactory agreement with experimental data
in the studies of the nuclei with deformed configuration 
and the giant resonances within the RPA 
formalism~\cite{HTT95,MT97,MA01}.
The hyperon masses are taken to be $m_{\Lambda}=1116 \ \rm{MeV}$,
$m_{\Sigma}=1193\ \rm{MeV}$, and $m_{\Xi}=1313\ \rm{MeV}$,
while the strange meson masses are $m_{\sigma^*}=975\ \rm{MeV}$ 
and $m_{\phi}=1020\ \rm{MeV}$~\cite{QMC99}.
As for the hyperon couplings, we employ the following relations 
derived from the quark model
\begin{equation}
\begin{array}{l}
\label{eq:relation}
 \frac{1}{3}g_{\sigma N}
=\frac{1}{2}g_{\sigma \Lambda}
=\frac{1}{2}g_{\sigma \Sigma}
=g_{\sigma \Xi}, 
 \\ [8pt] 
 \frac{1}{3}g_{\omega N}
=\frac{1}{2}g_{\omega \Lambda}
=\frac{1}{2}g_{\omega \Sigma}
=g_{\omega \Xi},
 \\[8pt]
 g_{\rho N}
=\frac{1}{2}g_{\rho \Sigma}
=g_{\rho \Xi}, \hspace{0.3cm} 
g_{\rho \Lambda}=0,
 \\[8pt]  
 2 g_{\sigma^* \Lambda}
=2 g_{\sigma^* \Sigma}
=g_{\sigma^* \Xi}
=\frac{2\sqrt{2}}{3}g_{\sigma N}, 
 \hspace{0.3cm} g_{\sigma^* N}=0,
 \\[8pt] 
 2 g_{\phi \Lambda}
=2 g_{\phi \Sigma}=g_{\phi \Xi}
=\frac{2\sqrt{2}}{3}g_{\omega N}, 
 \hspace{0.3cm} g_{\phi N}=0.
\end{array}
\end{equation}

In the RMF theory, the meson fields are treated as classical fields,
and the field operators are replaced by their expectation values.
The meson field equations in uniform matter are given by 
\begin{eqnarray}
\label{eq:m1}
& & m_\sigma^2\sigma-g_2 \sigma^2+g_3 \sigma^3 = 
\sum_B g_{\sigma B} \frac{2J_B + 1}{2\pi^2}
\int_0^{k_B} \frac{m_B^*}
{\sqrt{k^2 + m_B^{* 2}}} \: k^2 \ dk ,
\\
\label{eq:m2}
& & m_\omega^2\omega+c_3 \omega^3 = 
\sum_B g_{\omega B} \left(2J_B + 1\right)
k_B^3 \big/ (6\pi^2),
\\
\label{eq:m3}
& & m_\rho^2\rho = 
\sum_B g_{\rho B} I_{3B} \left(2J_B + 1\right)
k_B^3 \big/ (6\pi^2),
\\
\label{eq:m4}
& & m_{\sigma^*}^2\sigma^* = \sum_B g_{\sigma^* B} 
\frac{2J_B + 1}{2\pi^2} \int_0^{k_B} \frac{m_B^*}
{\sqrt{k^2 + m_B^{* 2}}} \: k^2 \ dk,
\\
\label{eq:m5}
& & m_\phi^2\phi = 
\sum_B g_{\phi B} \left(2J_B + 1\right)
k_B^3 \big/ (6\pi^2),
\end{eqnarray}
where $ m^*_B = m_B-g_{\sigma B}\sigma-g_{\sigma^* B}\sigma^*$
is the effective mass of the baryon species $B$, 
and $k_B$ is its Fermi momentum. 
$J_B$ and $I_{3B}$ denote the spin and the isospin projection
of baryon $B$.

For neutron star matter with uniform distributions, 
the composition is determined by the 
requirements of charge neutrality and  
$\beta$-equilibrium conditions. 
Considering the baryon octet and leptons included in the 
present calculation, the $\beta$-equilibrium conditions,
without trapped neutrinos, can be clearly expressed by  
\begin{eqnarray}
& & \mu_p=\mu_{\Sigma^+}=\mu_n-\mu_e, 
\nonumber\\
& & \mu_\Lambda=\mu_{\Sigma^0}=\mu_{\Xi^0}=\mu_n,
\nonumber\\
& & \mu_{\Sigma^-}=\mu_{\Xi^-}=\mu_n+\mu_e,
\nonumber\\
& & \mu_\mu=\mu_e,
\label{eq:beta}
\end{eqnarray}
where $\mu_i$ is the chemical potential of species $i$.
At zero temperature,
the chemical potentials of baryon $B$ and lepton $l$
are given by
\begin{eqnarray}
\label{eq:mu1}
& & \mu_B=\sqrt{k_B^2 + m_B^{* 2}} 
    +g_{\omega B}\omega + g_{\phi B}\phi + g_{\rho B}\tau_3\rho,
\\
\label{eq:mu2}
& & \mu_l=\sqrt{k_l^2 + m_l^{* 2}}. 
\end{eqnarray}
The charge neutrality condition has the following form:
\begin{equation}
\label{eq:charge}
 n_p+n_{\Sigma^+}=n_e+n_\mu+n_{\Sigma^-}+n_{\Xi^-},
\end{equation}
where $n_i=k_i^3 \big/ 3\pi^2$ is the number density of species $i$.
Then, the total baryon density is 
$n_B=n_n+n_p+n_\Lambda+n_{\Sigma^-}+n_{\Sigma^0}+n_{\Sigma^+}
    +n_{\Xi^-}+n_{\Xi^0}$.
We solve the coupled equations 
(\ref{eq:m1})-(\ref{eq:m5}), (\ref{eq:beta}), and (\ref{eq:charge})
self-consistently at a given baryon density $n_B$.
The total energy density and pressure of the uniform matter are given by
\begin{eqnarray}
\varepsilon &=& 
\sum_B \frac{2J_B +1}{2\pi^2} \int_0^{k_B}
\sqrt{k^2 + m_B^{* 2}} \ k^2  dk
+\frac{1}{2}m_\sigma^2 \sigma^2 
-\frac{1}{3} g_2\sigma^3 
+\frac{1}{4} g_3\sigma^4
\nonumber\\
 & &
+\frac{1}{2}m_\omega^2 \omega^2
+\frac{3}{4}c_3\omega^4
+\frac{1}{2} m_\rho^2 \rho^2
+\frac{1}{2}m_{\sigma^*}^2 \sigma^{* 2} 
+\frac{1}{2} m_\phi^2 \phi^2 
\nonumber\\
 & &
+ \sum_l \frac{1}{\pi^2} \int_0^{k_l} 
\sqrt{k^2 + m_l^2} \ k^2  dk \ ,
\label{eq:e}
\end{eqnarray}

\begin{eqnarray}
P &=& 
\frac{1}{3} \sum_B \frac{2J_B +1}{2\pi^2} \int_0^{k_B}
\frac{k^4 \ dk}{\sqrt{k^2 + m_B^{* 2}}} 
-\frac{1}{2}m_\sigma^2 \sigma^2 
+\frac{1}{3} g_2\sigma^3 
-\frac{1}{4} g_3\sigma^4
\nonumber\\
 & &
+\frac{1}{2}m_\omega^2 \omega^2
+\frac{1}{4}c_3\omega^4
+\frac{1}{2} m_\rho^2 \rho^2
-\frac{1}{2}m_{\sigma^*}^2 \sigma^{* 2} 
+\frac{1}{2} m_\phi^2 \phi^2 
\nonumber\\
 & &
+ \frac{1}{3} \sum_l \frac{1}{\pi^2} \int_0^{k_l} 
\frac{k^4 \ dk}{\sqrt{k^2 + m_l^2}} \ .
\label{eq:p}
\end{eqnarray}

\section{Thomas-Fermi approximation}

In the low density range, where heavy nuclei exist, 
we perform the Thomas-Fermi calculation based on the work done 
by Oyamatsu~\cite{O93}. 
In this approximation, the non-uniform matter 
can be modeled as a lattice of nuclei 
immersed in a vapor of neutrons and electrons. 
At lower density, there is no neutron dripping out of nuclei. 
We assume that each heavy spherical nucleus is
located in the center of a charge-neutral cell consisting of 
a vapor of neutrons and electrons. The nuclei form a
body-centered-cubic (BCC) lattice to minimize the Coulomb 
lattice energy. It is useful to introduce the Wigner-Seitz 
cell to simplify the energy of the unit cell. 
The Wigner-Seitz cell is a sphere whose volume is
the same as the unit cell in the BCC lattice.

We assume the nucleon distribution functions $n_i(r)$
($i=n$ for neutron, $i=p$ for proton)
in the Wigner-Seitz cell as
\begin{equation}
n_i\left(r\right)=\left\{
\begin{array}{ll}
\left(n_i^{in}-n_i^{out}\right) \left[1-\left(\frac{r}{R_i}\right)^{t_i}
\right]^3 +n_i^{out},  & 0 \leq r \leq R_i \\
n_i^{out},  & R_i \leq r \leq R_{cell} \\
\end{array} \right. ,
\label{eq:dis}
\end{equation}
where $r$ represents the distance from the center of the nucleus,
and $R_{cell}$ is the radius of the Wigner-Seitz cell 
defined by the relation,
$V_{cell}=\frac{4\pi}{3} R_{cell}^3 =a^3$ ($a$ is the lattice constant).
The parameters $R_i$ and $t_i$ determine the boundary and the relative surface thickness of the heavy nucleus.
$R_n$ and $t_n$ may be a little different from $R_p$ and $t_p$ due to 
the additional neutrons forming a neutron skin in the surface region.
For neutron star matter at a given average density of baryons,
$n_B=\int_{cell}\left[ n_n\left(r\right)
    + n_p\left(r\right)  \right] d^3r \ \big/ \ V_{cell} $, 
there are only seven independent parameters among the eight variables: 
$a, n_n^{in}, n_n^{out}, R_n, t_n, n_p^{in}, R_p, t_p$.
The optimal state is determined by minimizing the average energy density,
$\varepsilon= E_{cell}\ \big/ \ V_{cell} $,
with respect to those independent parameters.

The total energy per cell, $E_{cell}$, can be written as 
\begin{equation}
E_{cell}=E_{bulk}+E_s+E_C+E_e \ .
\label{eq:etot}
\end{equation}
Here the bulk energy of baryons, $E_{bulk}$, is calculated by
\begin{equation}
E_{bulk}=\int_{cell} \varepsilon_{RMF} 
         \left( \ n_n\left(r\right), n_p\left(r\right) \ \right) d^3r,
\end{equation}
where $\varepsilon_{RMF}$ is the energy density in the RMF theory
as the functional of the neutron density $n_n$ and the proton
density $n_p$. As the input in the Thomas-Fermi calculation, 
$\varepsilon_{RMF}$ at each radius $r$ is calculated
in the RMF theory for uniform matter with the corresponding 
densities $n_n$ and $n_p$.
The surface energy term $E_s$ due to the inhomogeneity of nucleon
distribution is given by,
\begin{equation}
E_s=\int_{cell} F_0 \mid \nabla \left( \ n_n\left(r\right)+
    n_p\left(r\right) \ \right) \mid^2 \ d^3r,
\end{equation}
where the parameter $F_0=70 \ \rm{MeV\cdot fm^5}$ 
is determined by doing the Thomas-Fermi calculations of finite nuclei  
as described in the appendix in Ref.~\cite{O93}. 
The Coulomb energy per cell $E_C$ is calculated using the Wigner-Seitz 
approximation with an added correction term for the BCC lattice:
\begin{equation}
E_C=\frac{1}{2}\int_{cell} e\  
    \left[n_p\left(r\right)-n_e\right]\ \phi(r) \ d^3r
    \ +\ \triangle E_C,
\end{equation}
where $\phi(r)$ stands for the electrostatic potential calculated 
in the Wigner-Seitz approximation,
$\triangle E_C=C_{BCC} (Z e)^2\big/a$ is the correction term 
for the BCC lattice as given in Ref.~\cite{O93}.
$n_e$ is the electron number density of uniform electron gas,
which can be determined by the charge neutrality condition as
$ n_e= Z \big/ V_{cell} $ ($Z$ denotes the proton number per cell).
The last term in Eq. (\ref{eq:etot}), $E_e$, 
is the kinetic energy of electrons, which is given by
\begin{equation}
E_e=\frac{1}{\pi^2} \int_0^{k_e} \sqrt{k^2 + m_e^2} \ k^2  dk \,
\end{equation}
where $k_e=(3\pi^2 n_e)^{1/3}$ is the Fermi momentum of electrons.

For each baryon density $n_B$, we minimize the average energy density 
$\varepsilon$ of non-uniform matter
with respect to the independent parameters in the Thomas-Fermi
approximation. At some higher densities, the heavy nuclei dissolve 
and the matter becomes homogeneous. We determine the density, at which
the phase transition takes place, by comparing the energy density 
of non-uniform matter with  the one of uniform matter.

\section{Properties of neutron star matter}

\begin{figure}[htb]
\vspace{0.1cm}
\epsfxsize=12cm
\epsfysize=16cm
\centerline{\epsfbox{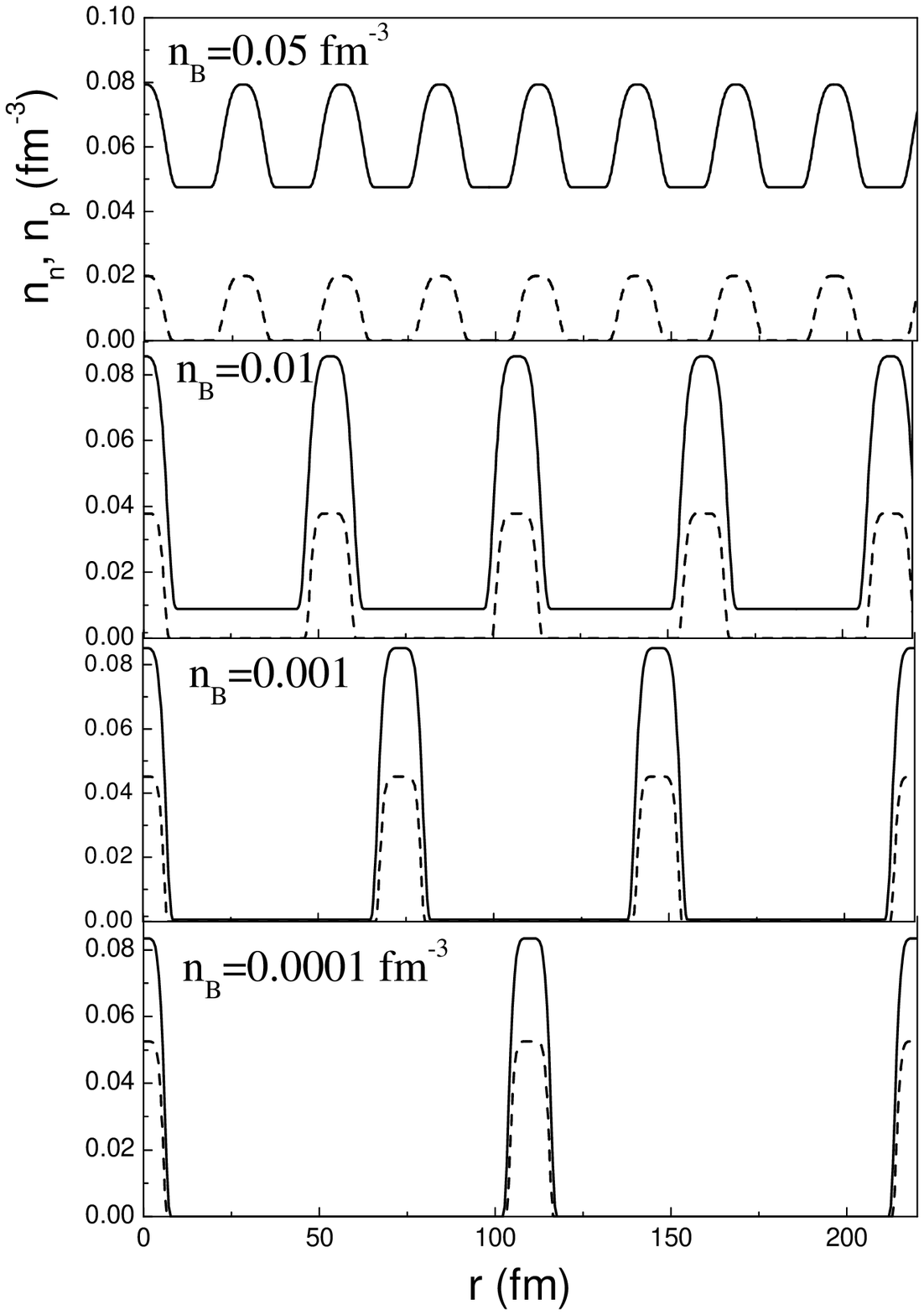}}
\caption{
The neutron distribution (solid curves) and the proton distribution 
(dashed curves) along the straight lines joining the centers of
the nearest nuclei in the BCC lattice at the average baryon density
$n_B = 0.0001, \ 0.001, \ 0.01, \ 0.05\ \rm{fm^{-3}}$.
}
\label{fig:dis}
\end{figure}

\begin{figure}[htb]
\vspace{0.1cm}
\epsfxsize=12cm
\epsfysize=16cm
\centerline{\epsfbox{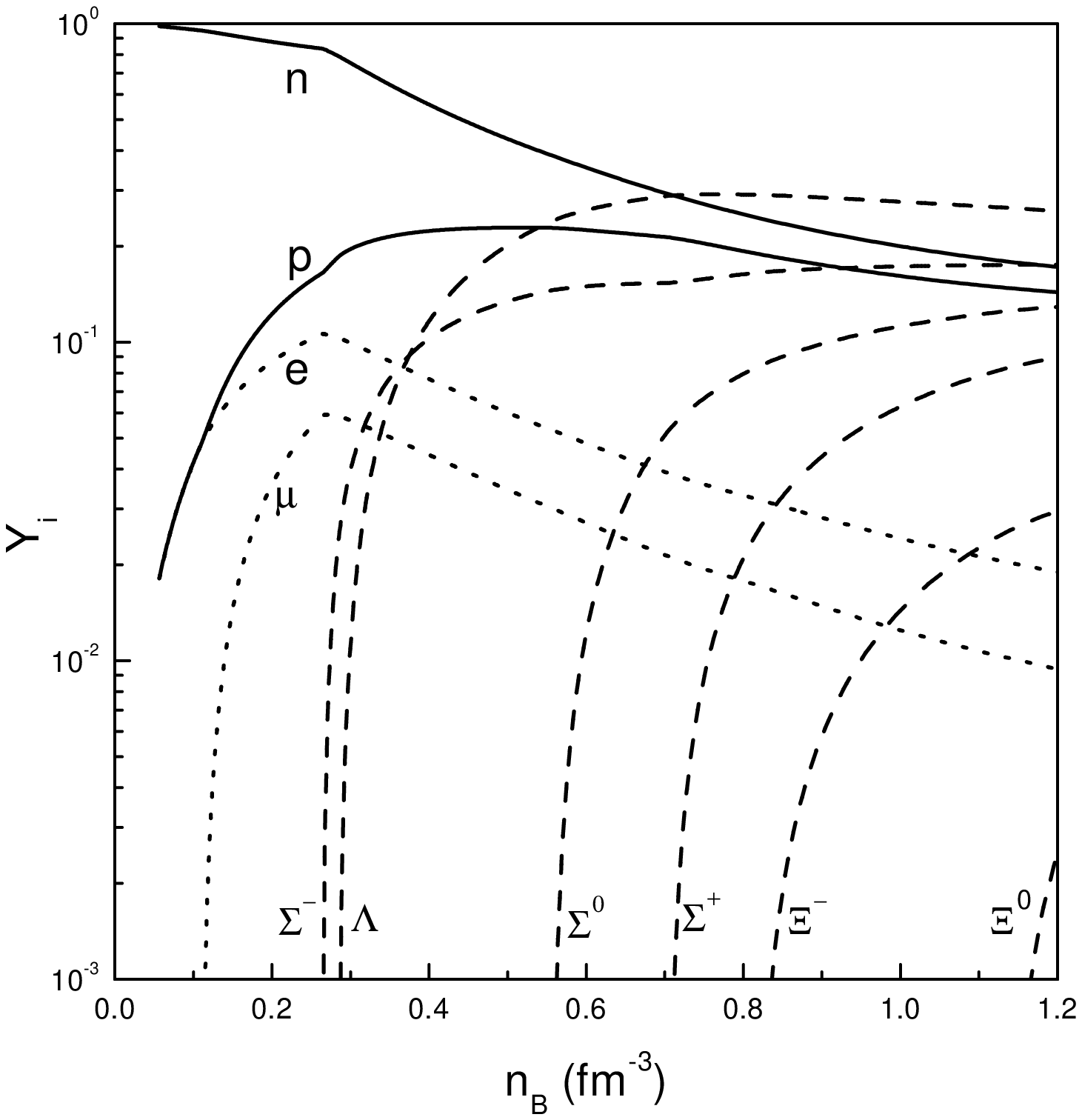}}
\caption{
The fraction of species $i$, $Y_i=n_i\big/n_B$,
as a function of the total baryon density $n_B$.
}
\label{fig:eosy}
\end{figure}

\begin{figure}[htb]
\vspace{0.1cm}
\epsfxsize=12cm
\epsfysize=16cm
\centerline{\epsfbox{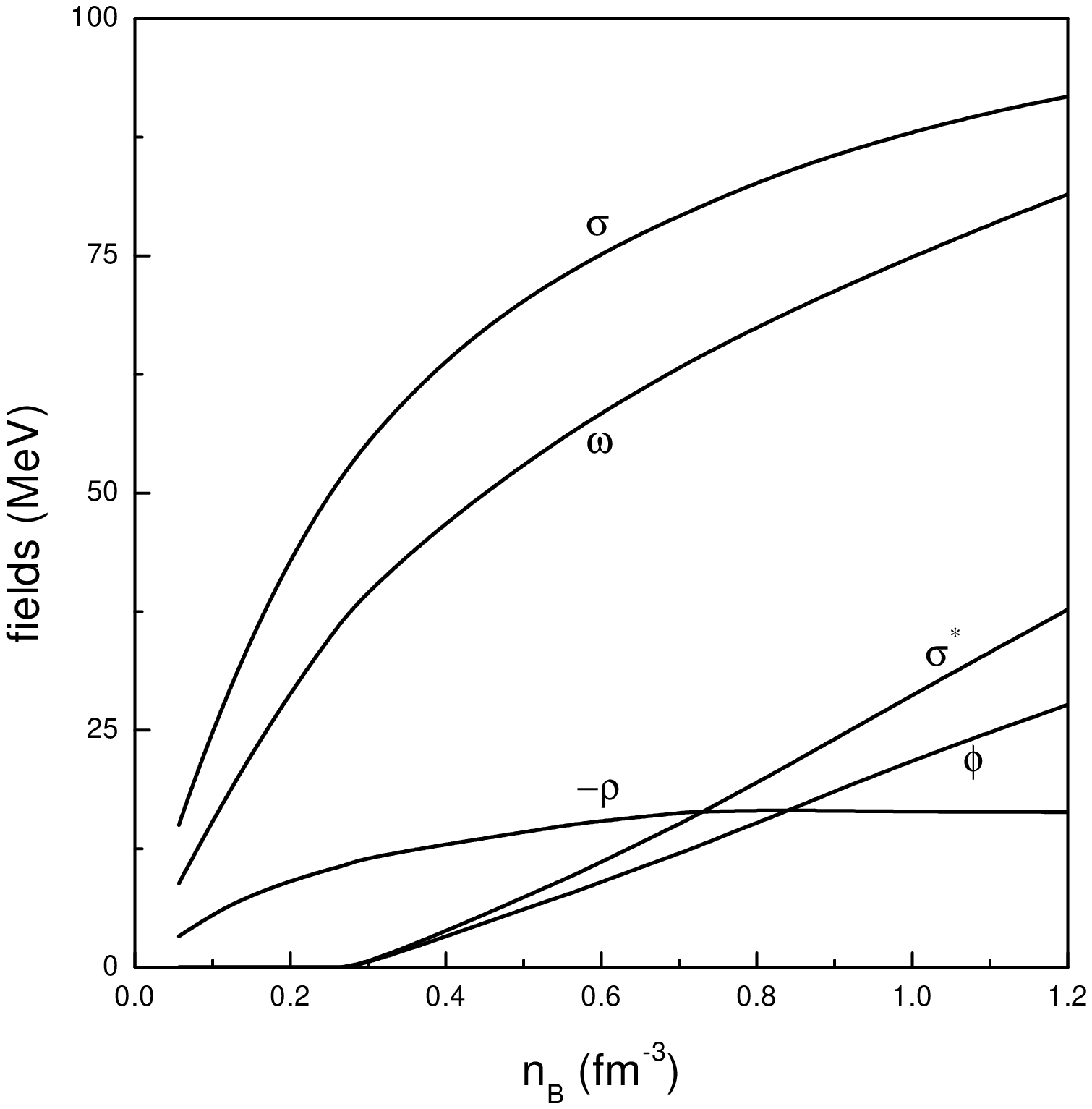}}
\caption{
The meson mean fields as functions of baryon density.
}
\label{fig:field}
\end{figure}

In this section, we present the resulting EOS of neutron star 
matter in the density range
from $10^{-7}$ to $1.2\ \rm{fm^{-3}}$.
At low densities where heavy nuclei exist, 
the non-uniform matter is described by the Thomas-Fermi 
approximation, in which the optimal state is determined 
by minimizing the average energy density with respect to its 
independent parameters.
For the density below $\sim 2.4\times 10^{-4}\ \rm{fm^{-3}}$,
the nucleons form the optimal nuclei and those nuclei 
build a BCC lattice with uniform electron gas.
It is found that the neutrons begin to drip out from nuclei 
at $n_B \sim 2.4\times 10^{-4}\ \rm{fm^{-3}}$, then there is 
a neutron gas in addition to the electron gas.
In Fig. \ref{fig:dis} we show the neutron and the proton 
distributions along the straight line joining the centers
of the nearest nuclei in the BCC lattice at the average baryon 
densities $n_B = 0.0001, \ 0.001, \ 0.01, \ 0.05\ \rm{fm^{-3}}$.
As the density increases, the optimal nuclei become closer and 
more neutron rich. 
At $n_B \sim 0.06\ \rm{fm^{-3}}$, the nuclei dissolve 
and the optimal state is a uniform matter 
consisting of neutrons, protons, and electrons in $\beta$-equilibrium.
When the electron chemical potential exceeds the rest mass
of the muon (at $n_B \approx 0.11\ \rm{fm^{-3}}$),  
it becomes energetically favorable to convert the electrons at the 
Fermi surface into muons, then the muons appear with the chemical
equilibrium condition $\mu_e=\mu_\mu$.
Hyperons appear at higher densities 
($n_B \stackrel{>}{\sim} 0.27\ \rm{fm^{-3}}$).
In Fig. \ref{fig:eosy} we show the fraction of species 
$i$, $Y_i=n_i\big/n_B$, as a function of the total baryon density $n_B$. 
The composition of uniform neutron star matter is calculated by 
solving the coupled equations (\ref{eq:m1})-(\ref{eq:m5}), 
(\ref{eq:beta}), and (\ref{eq:charge}).
The threshold density for a hyperon species is determined 
not only by its charge and mass but also by the meson mean fields,
which are shown in Fig. \ref{fig:field} as functions of baryon density. 
In the present calculation, $\Sigma^-$ is the first hyperon which appears
at $n_B \approx 0.27\ \rm{fm^{-3}}$, while $\Lambda$ has almost the same 
threshold density ($n_B \approx 0.29\ \rm{fm^{-3}}$).
It is partly because the negative charge is much more favorable, 
even though $\Sigma^-$ has somewhat larger mass compared with the mass
of $\Lambda$. The other hyperons, $\Sigma^0$, $\Sigma^+$, $\Xi^-$,
and $\Xi^0$, appear one by one at higher densities  
($n_B \approx 0.57, \ 0.72, \ 0.84, \ 1.17\ \rm{fm^{-3}}$).
The appearance of hyperons causes some decreases 
of the nucleon fractions.
At high densities ($n_B \stackrel{>}{\sim} 0.7\ \rm{fm^{-3}}$), 
the $\Lambda$ fraction is larger than the neutron fraction.
We note that the hyperon threshold densities and fractions are 
sensitive to the hyperon couplings, 
and there are quite large uncertainties in the hyperon couplings.
In this work, we adopt the hyperon couplings derived from the quark model.

\begin{figure}[htb]
\vspace{0.1cm}
\epsfxsize=12cm
\epsfysize=16cm
\centerline{\epsfbox{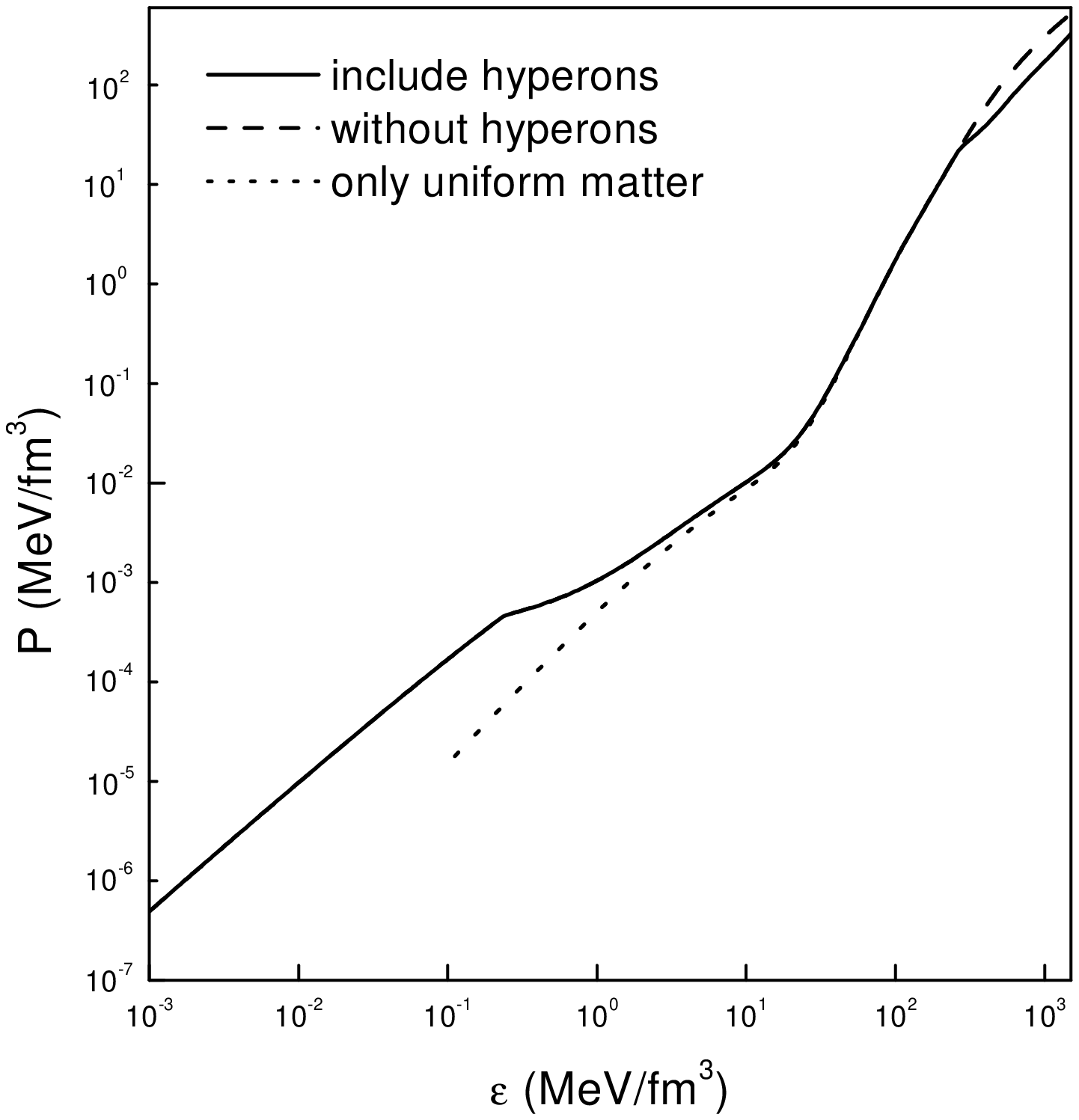}}
\caption{
The pressure $P$ versus energy density $\varepsilon$ 
for neutron star matter with the inclusion of hyperons (solid curve)
and without hyperons (dashed curve).  The EOS considering only
uniform matter phase (dotted curve) is also shown for comparison.
}
\label{fig:eosep}
\end{figure}

\begin{figure}[htb]
\vspace{0.1cm}
\epsfxsize=12cm
\epsfysize=16cm
\centerline{\epsfbox{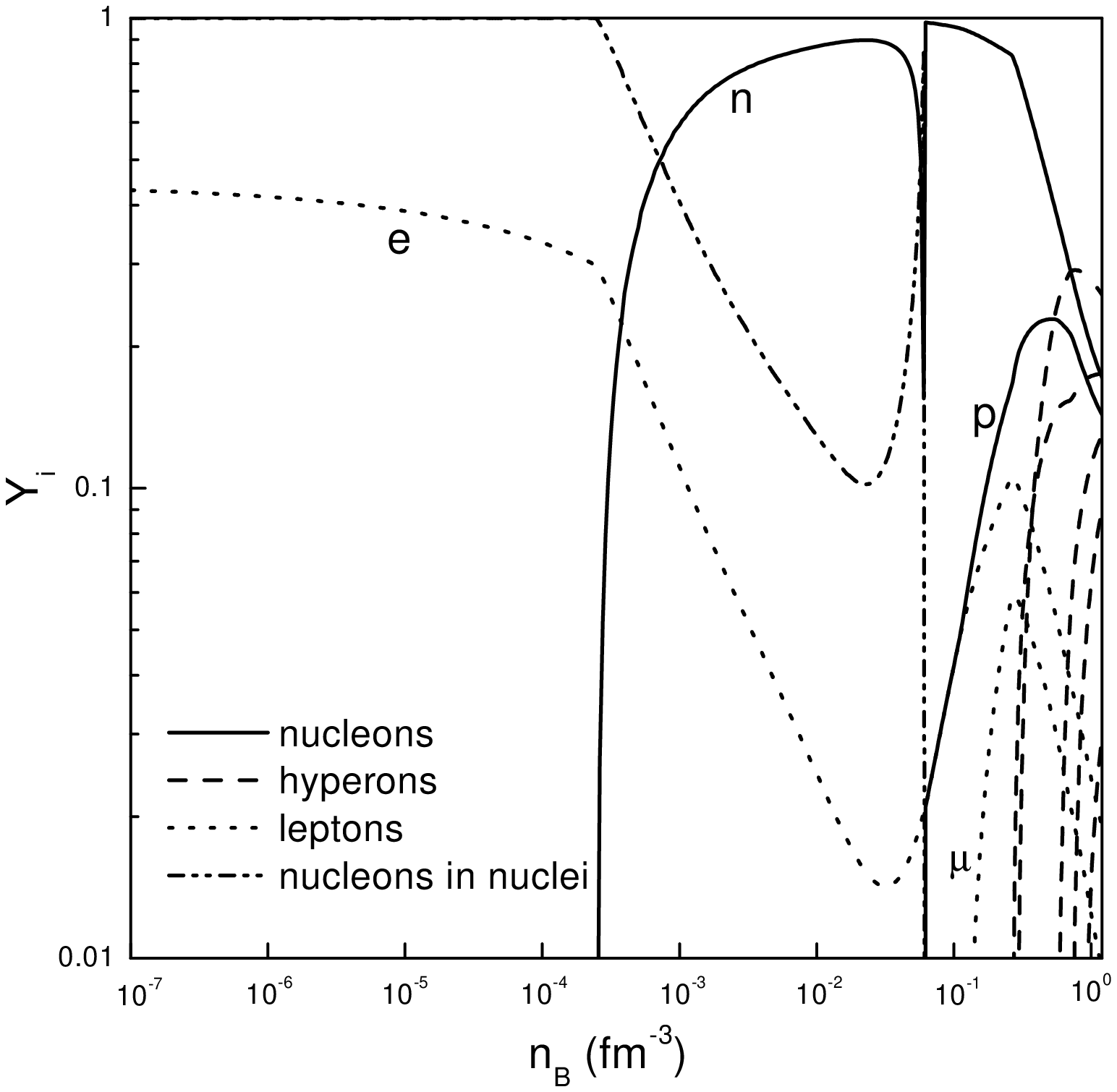}}
\caption{The fractions of the composition in neutron star
         matter as functions of baryon density.} 
\label{fig:tab}
\end{figure}

We display in Fig. \ref{fig:eosep} the pressure of neutron star matter
as a function of energy density. The present EOS shown 
by the solid curve is compared with the EOS considering only
the uniform matter phase (dotted curve), and it is found that 
the contribution from the non-uniform matter is quite large 
at low densities. The EOS without hyperons is also shown for comparison
by dashed curve. The inclusion of hyperons considerably softens
the EOS at high densities, because the conversion of nucleons 
to hyperons can relieve the Fermi pressure of the nucleons.
In Fig. \ref{fig:tab} we show the fraction of species $i$, $Y_i$,
in neutron star matter as a function of the average baryon density $n_B$.
It is very interesting to see the phase transitions in the wide density
range. At low densities, all nucleons exist inside nuclei, therefore
the fraction of the nucleons in nuclei (dot-dashed curve) is equal to one.
The decrease of the electron fraction (dotted curve), 
which is equal to the proton fraction due to the charge neutrality, 
implies that the optimal nucleus becomes more and more neutron rich 
as the density increases. Beyond the neutron drip density
($n_B \sim 2.4\times 10^{-4}\ \rm{fm^{-3}}$), there is a 
increasing fraction of the free neutrons outside nuclei (solid curve),
and this causes a rapid decrease of the fraction of the nucleons
in nuclei (dot-dashed curve).
The phase transition from non-uniform matter to uniform matter
occurs at $\sim 0.06\ \rm{fm^{-3}}$, where the heavy nuclei dissolve
and the matter consists of neutrons, protons, and electrons
in $\beta$-equilibrium. We note that the neutron star matter is 
assumed to be at zero temperature, so there is no free proton gas 
outside nuclei in the non-uniform matter phase. 
The muon fraction appears at $n_B \approx 0.11\ \rm{fm^{-3}}$
with the charge neutrality condition $Y_\mu+Y_e=Y_p$.
At high densities ($n_B \stackrel{>}{\sim} 0.27\ \rm{fm^{-3}}$), 
the hyperon fractions appear, 
which have been shown more clearly in Fig. \ref{fig:eosy}.    

\section{Neutron star structure}

\begin{figure}[htb]
\vspace{0.1cm}
\epsfxsize=12cm
\epsfysize=16cm
\centerline{\epsfbox{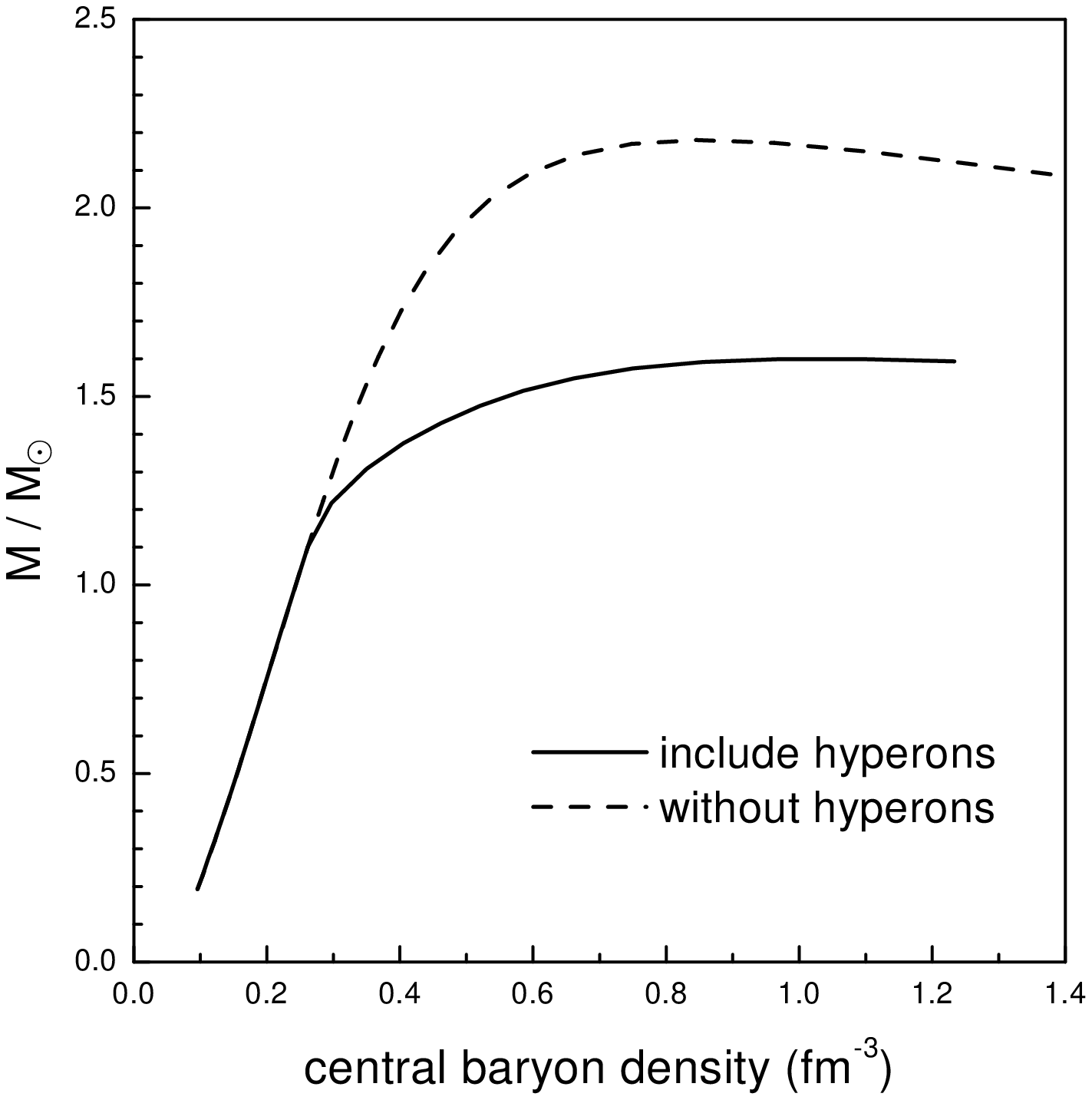} }
\caption{The neutron star masses as functions of central baryon
         density}
\label{fig:nstarmass}
\end{figure}

\begin{figure}[htb]
\vspace{0.1cm}
\epsfxsize=12cm
\epsfysize=16cm
\centerline{\epsfbox{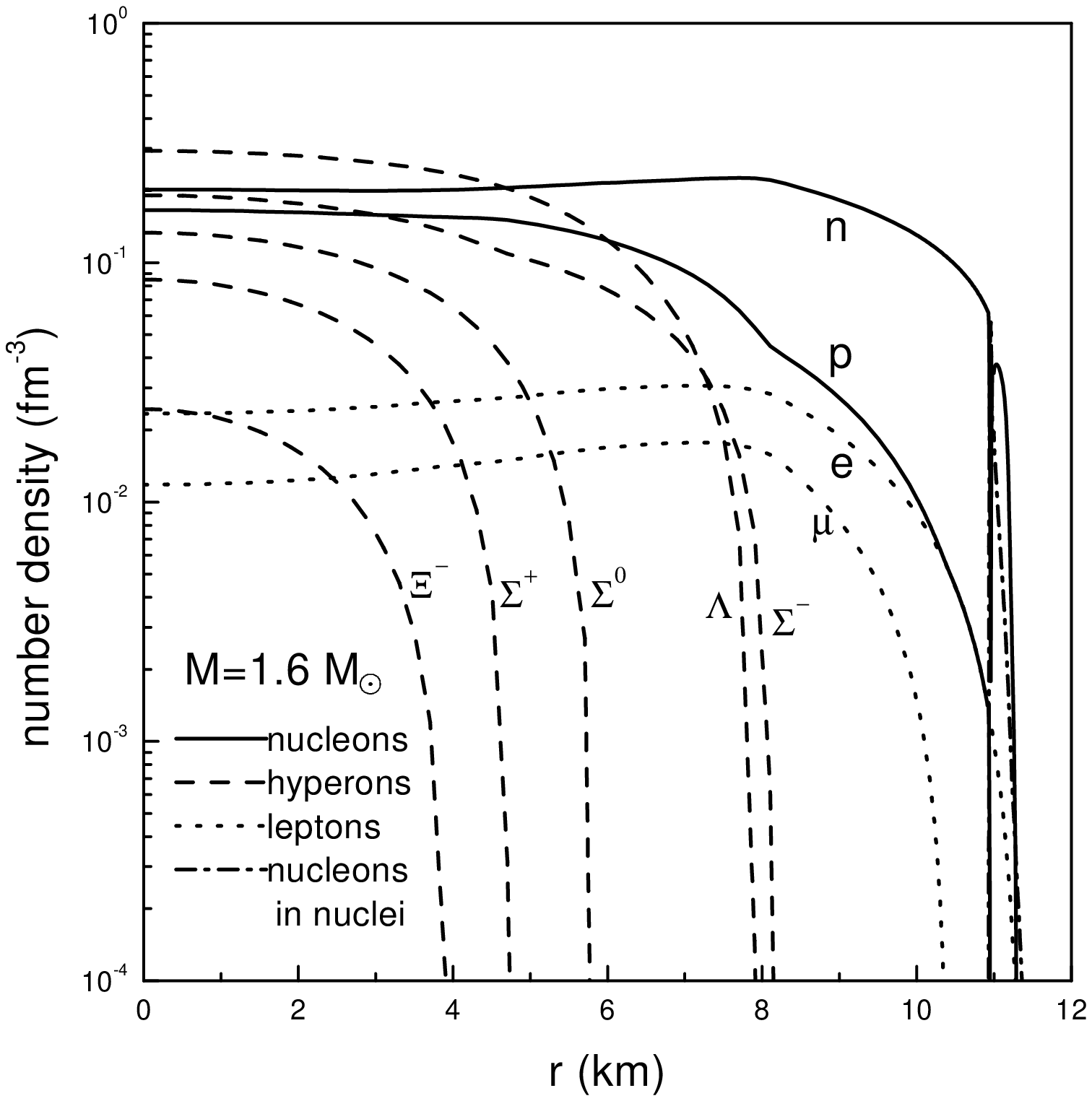} }
\caption{The number density of the composition in the 
         neutron star with $M=1.6 M_\odot$ 
         as a function of radius $r$.}
\label{fig:nstary16}
\end{figure}

\begin{figure}[htb]
\vspace{0.1cm}
\epsfxsize=12cm
\epsfysize=16cm
\centerline{\epsfbox{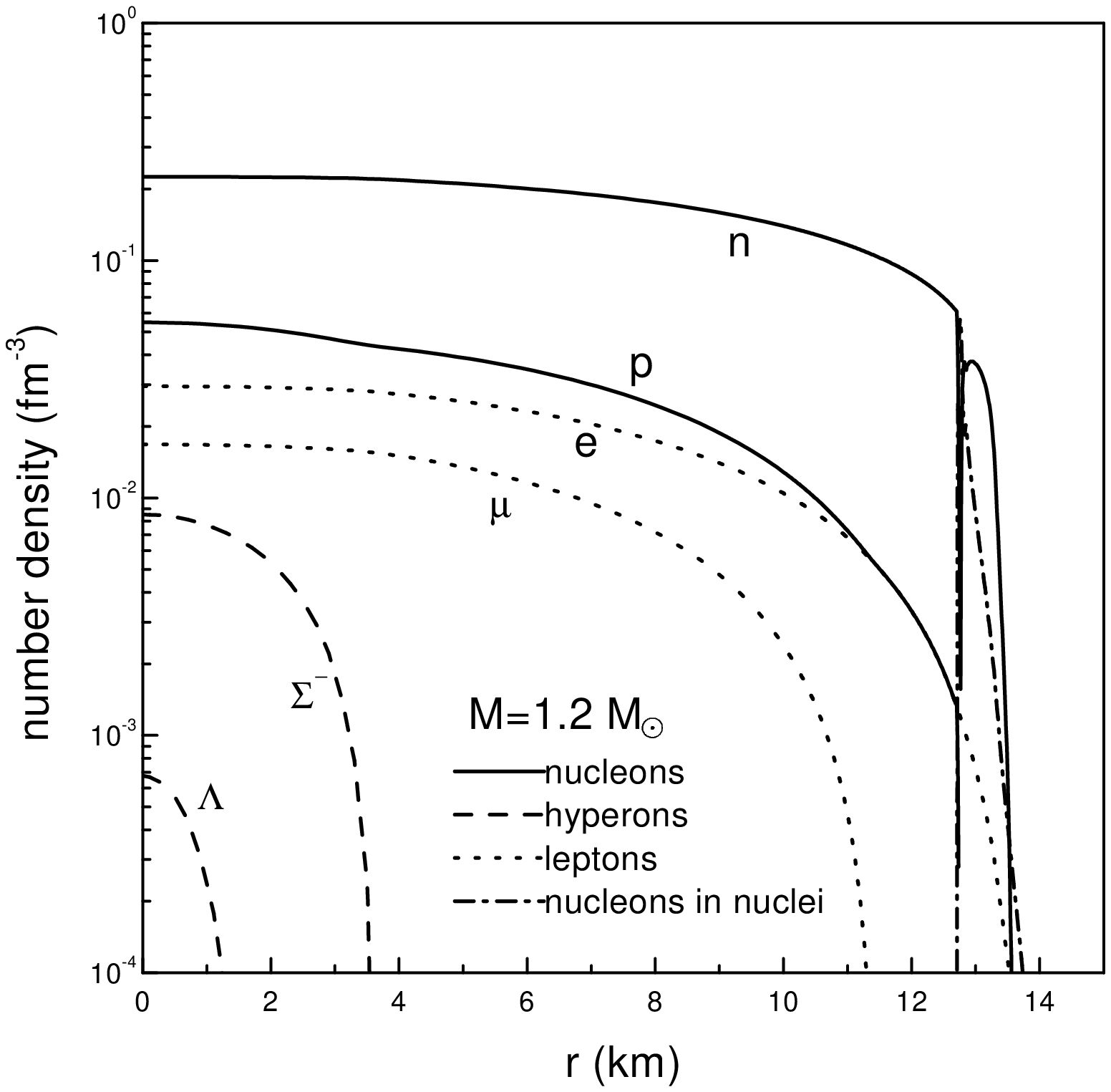} }
\caption{Same as Fig. \ref{fig:nstary16} but for $M=1.2 M_\odot$.}
\label{fig:nstary12}
\end{figure}

\begin{figure}[htb]
\vspace{0.1cm}
\epsfxsize=12cm
\epsfysize=16cm
\centerline{\epsfbox{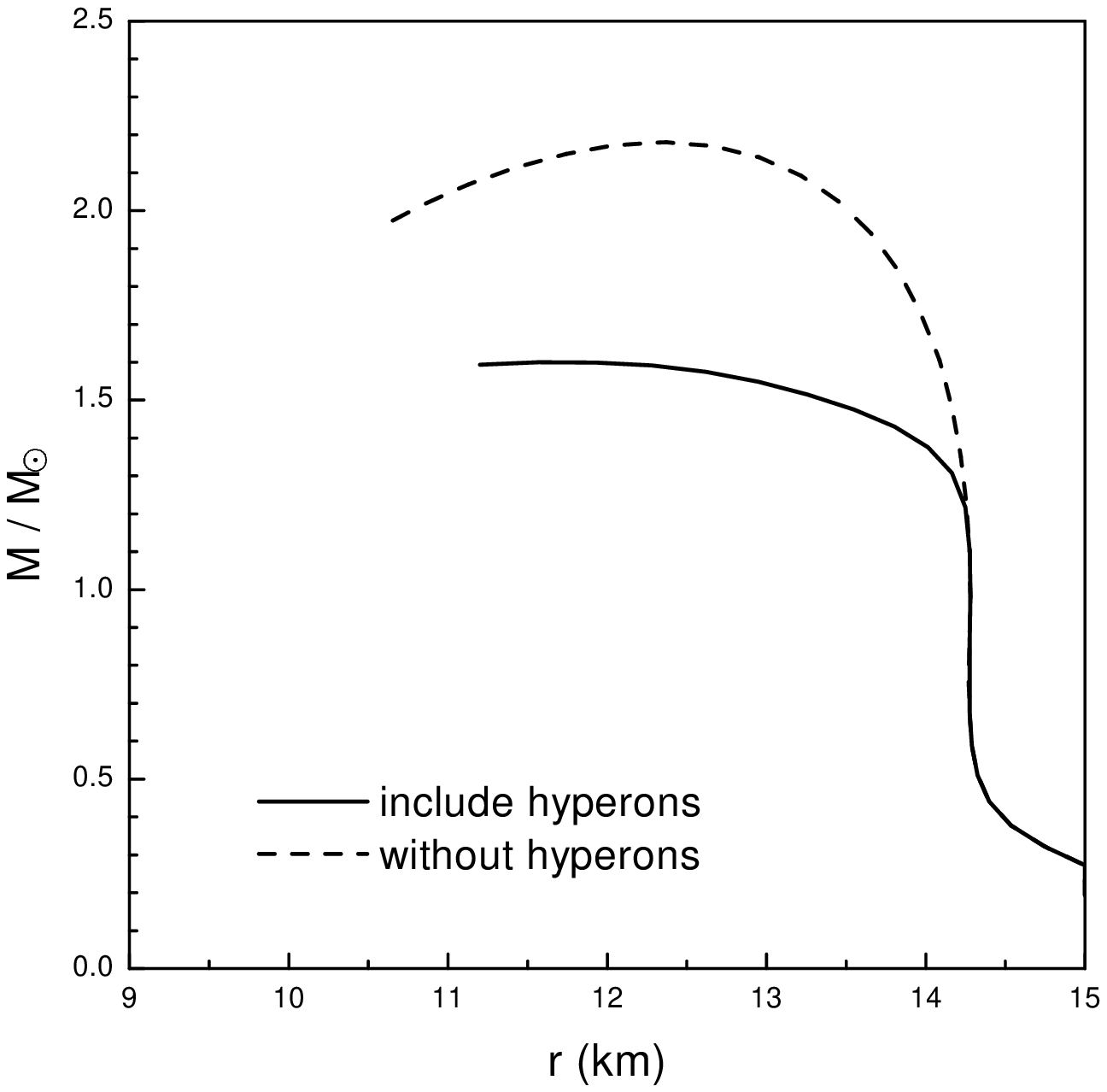} }
\caption{The mass-radius relations for neutron stars. 
         The solid curve shows the results with the inclusion of 
         hyperons, while those without hyperons is plotted by
         the dashed curve for comparison.}
\label{fig:nstarmr}
\end{figure}

We calculate the neutron star properties by using the relativistic EOS.
The neutron star masses as functions of central baryon density
are displayed in Fig. \ref{fig:nstarmass}. 
It is shown that the maximum mass of the neutron stars including 
hyperons is around $1.6 M_\odot$,
while it is around $2.2 M_\odot$ without hyperons.
The neutron star mass is determined predominantly by the behavior of 
the EOS at high densities. The inclusion of hyperons considerably 
softens the EOS at high densities, therefore, results much 
smaller neutron star masses.
The non-uniform matter, which exists in the crusts of neutron stars,
has negligible contribution to the total neutron star mass, 
but it plays an important role in the description of 
the neutron star profile in the crustal region.
In Fig. \ref{fig:nstary16} and \ref{fig:nstary12}, 
we show the number density of the composition in the 
neutron stars with $M=1.6 M_\odot$ and $M=1.2 M_\odot$ respectively,
as a function of radius. 
It is clear that the uniform matter containing the equilibrium 
mixture of nucleons, hyperons, and leptons exists in the internal 
region of the neutron star, while the non-uniform matter phase 
occurs only in the surface region. 
The neutron star with $M=1.6 M_\odot$ has much thinner crusts 
as compared to the case of the neutron star with $M=1.2 M_\odot$.
We show in Fig. \ref{fig:nstarmr} the mass-radius relations
using the EOS with or without hyperons.
It is found that the inclusion of hyperons only influences 
the neutron stars having large masses 
($M \stackrel{>}{\sim} 1.2 M_\odot$).

\section{CONCLUSION}

We have constructed the relativistic EOS of neutron star matter 
in the density range from $10^{-7}$ to $1.2\ \rm{fm^{-3}}$.
The non-uniform matter at low densities has been described 
by the Thomas-Fermi approximation, in which the nucleons form 
the optimal nuclei and those nuclei build a BCC lattice.
The uniform matter at high densities has been studied in the RMF theory.
We adopted the RMF model with the TM1 parameter set,
which was demonstrated to be successful in describing the 
properties of nuclear matter and finite nuclei including 
unstable nuclei~\cite{ST94}, and its results were taken as  
the input in the Thomas-Fermi calculations. 
Hence we have worked out consistent calculations 
for uniform matter and non-uniform matter.
The phase transition from non-uniform matter to uniform matter
is found to take place at $n_B \sim 0.06 \ \rm{fm^{-3}}$.
At high densities ($n_B \stackrel{>}{\sim} 0.27\ \rm{fm^{-3}}$),
it is energetically favorable to convert some nucleons into hyperons 
via weak interactions.
The inclusion of hyperons leads to a considerable softening of 
the EOS at high densities, since the conversion of nucleons 
to hyperons can relieve the Fermi pressure of the nucleons.
We note that the contributions from hyperons are sensitive 
to the hyperon couplings, here we have adopted the hyperon couplings 
derived from the quark model.
Presently, there exist large uncertainties in hyperon couplings.
The hyperon couplings should be constrained by
the experimental data of hypernuclei,
but the experimental information is deficient for determining them.
From the study of single $\Lambda$ hypernuclei,  
the quark model values of $\Lambda$ hyperon couplings 
usually predict overbinding of $\Lambda$ single particle energies.
It seems that the quark model values of $\Lambda$ hyperon couplings
lead to rather strong attraction. 
This might cause an earlier appearance of $\Lambda$ hyperon.  
A detailed investigation of the dependence of the results on
the hyperon couplings is deferred to future work.

We have employed the present EOS to calculate the neutron star properties.
With the appearance of hyperons, the maximum mass of neutron stars 
turned out to be $1.6 M_\odot$. It is found that the inclusion of 
hyperons results much smaller neutron star masses due to the softening
of the EOS. The core of massive neutron stars is then composed of the 
equilibrium mixture of nucleons, hyperons, and leptons.
The non-uniform matter exists only in the surface region,
which forms quite thin crusts of neutron stars.  
The consideration of the non-uniform matter phase has negligible
contribution to the neutron star mass, but it is essential
to provide a realistic description of the neutron star structure.

The present calculations have been performed within the
framework of the relativistic mean field approach,
which is incapable to include pions explicitly.
It will be possible and important to construct a complete
EOS based on more microscopic theory such as 
the Dirac-Brueckner-Hartree-Fock approach.
Especially, same approach should be employed in the treatment
of both uniform matter and non-uniform matter.
It is well known that the relativity plays an essential role 
in describing the nuclear saturation and the nuclear structure, 
it also brings some distinctive properties in the EOS comparing 
with the case in the non-relativistic framework.
Therefore, it is very interesting and important to study the 
astrophysical phenomena such as neutron star properties 
using the relativistic EOS.

\section*{Acknowledgments}
The author would like to thank H. Toki, K. Sumiyoshi, and K. Oyamatsu
for fruitful discussions and collaborations.
This work was supported in part by the National Natural Science
Foundation of China under contract No. 10075028 and No. 10135030.


\end{document}